\documentclass[copyright,creativecommons]{eptcs}
\providecommand{\event}{GASCom 2024} 

\usepackage{iftex}

\newtheorem{proposition}{Proposition}

\ifpdf
  \usepackage{underscore}         
  \usepackage[T1]{fontenc}        
\else
  \usepackage{breakurl}           
\fi

\title{Detecting Isohedral Polyforms with a SAT Solver}
\author{Craig S. Kaplan
\institute{School of Computer Science \\ University of Waterloo}
\email{csk@uwaterloo.ca}
}
\def\titlerunning{Detecting Isohedral Polyforms with a SAT Solver}
\def\authorrunning{C.S. Kaplan}
\begin{document}
\maketitle

\begin{abstract}
I show how to express the question of whether a polyform tiles
the plane isohedrally as a Boolean formula that can be tested using
a SAT solver.  This approach is adaptable to a wide range of
polyforms, requires no special-case code for different isohedral
tiling types, and integrates seamlessly with existing software for
computing Heesch numbers of polyforms.
\end{abstract}

\section{Introduction}

The study of algorithms for computing tiling-theoretic properties
of shapes is a rich and fascinating branch of computational geometry.
Implementations of these algorithms can also serve as useful tools
in the experimental side of tiling theory, as part of the search
for new shapes with interesting properties.  For example, Myers
systematically computed isohedral numbers (the minimum number of
transitivity classes in any tiling by a given shape) for many simple
polyforms~\cite{Myers}.  Building on Myers's work, I computed Heesch
numbers (the maximum number of times that a non-tiling shape can
be surrounded by layers of copies of itself) for simple polyforms~\cite{Kaplan}.
Our tools did not contribute to Smith's initial discovery of the
``hat'' aperiodic monotile, but they played a central
role in our subsequent analysis of the hat and our proof (with
Goodman-Strauss) of its aperiodicity~\cite{Hat}.

An \emph{isohedral tiling} is a tiling by congruent copies of some
\emph{prototile} $T$, such that for any two tiles $T_1$ and $T_2$
there exists a symmetry of the tiling mapping $T_1$ to $T_2$.
Isohedral tilings are some of the simplest periodic tilings, in that
all tiles belong to a single transitivity class relative to the 
symmetries of the tiling.  A complete theory of isohedral tilings,
including their classification into 81 tiling types with unmarked
tiles, was worked out by Gr\"unbaum and Shephard~\cite[Chapter 6]{GS}.

Given a simple shape such as a polyform, does it admit any isohedral
tilings?  This question offers interesting opportunities for the
development of new algorithms. It is also of practical interest as
part of any software for computing the tiling-theoretic properties
of shapes.  Myers's software~\cite{Myers} can detect isohedral
prototiles quickly, but formal questions of computational complexity
are more or less peripheral to his work.  The current state of the
art, at least for the special case of polyominoes, is the quasilinear-time 
algorithm by Langerman and Winslow~\cite{WL}.

In this paper I present a new technique for checking whether a polyform
tiles isohedrally. The algorithm is based on expressing the question
as a Boolean formula that can be checked by a SAT solver, and was motivated
by my desire to integrate such a test into my existing SAT-based framework
for computing Heesch numbers~\cite{Kaplan}.  I will explain the 
mathematical basis for this approach (Section~\ref{sec:math}), 
followed by its expression in Boolean logic (Section~\ref{sec:comp}),
and then conclude with a few final observations (Section~\ref{sec:conclusion}).

\section{Identifying prototiles based on surrounds}
\label{sec:math}

In order to determine whether a shape
admits any isohedral tilings of the plane, it suffices to examine
the ways that the shape can be surrounded by copies of itself. 
That is, if there exists a surround with a particular structure
that will be explained here, then the shape is guaranteed to tile
isohedrally. 

Let $T$ be a shape, which in full generality can be any topological
disk, but which for my purposes is typically a polygon.  Without
loss of generality, I assume here that $T$ is asymmetric.  (A
symmetric shape can always be decorated with an asymmetric marking,
with the meaning of congruence expanded to preserve markings.)

A \emph{patch} is a finite collection of congruent copies of $T$,
with pairwise disjoint interiors, whose union is a topological disk.
In particular, if exactly one copy of $T$ lies in the interior of
the patch, then we refer to the patch as a \emph{$1$-patch}, to the
interior tile as the patch's \emph{centre}, and to the remaining
tiles as a \emph{surround} of $T$.

The fact that every two tiles in an isohedral tiling are related
by a symmetry of the tiling implies that every tile is the centre
of a congruent $1$-patch, or more loosely that tiles have congruent
surrounds.  Gr\"unbaum and Shephard use this fact to develop a
complete enumeration of isohedral tiling types, based on an ``incidence
symbol'' that expresses a prototile's relationships to its
neighbours~\cite{GS}.  In fact, the converse holds as well: Dolbilin
and Schattschneider showed that if the tiles in a tiling have
congruent surrounds, then the tiling must be isohedral~\cite{Local}.

Let $\mathcal{S}=\{T_1,\ldots,T_n\}$ be a surround of a shape
$T$.  The surround is made up of congruent copies of $T$,
meaning that each $T_i=g_i(T)$ for some rigid motion $g_i$.
Fix one shape $T_i$ in the surround, and construct 
$\mathcal{S}_i=\{g_i\circ g_j(T)\}_{j=1}^n$, a congruent copy 
of $\mathcal{S}$ placed around $T_i$.  I call $T_i$ \emph{extendable}
if this transformed surround does not ``conflict'' with $T_i$'s neighbours 
in the original $1$-patch centred at $T$.  More precisely, $T_i$ is
extendable if for every 
$A\in\{T,T_1,\ldots,T_n\}$ and every $B\in\mathcal{S}_i$, either
$A=B$ or $A$ and $B$ have disjoint interiors.

Suppose that $T$ has a surround in which every $T_i$ is extendable.
The transformed surrounds $\mathcal{S}_i$ must all be compatible
with the $1$-patch around~$T$ and with each other, meaning that
their union will surround $\mathcal{S}$ with a second layer of
tiles.  In this manner we can continue outward layer by layer, each
time completing the surrounds of the tiles along the boundary of
the growing patch.  (This construction is similar to one used by
Gr\"unbaum and Shephard~\cite[Theorem 6.1.1]{GS}.)
In the limit we obtain a tiling of the plane
in which every tile has a congruent surround, which must therefore
be isohedral by The Local Theorem of Dolbilin and
Schattschneider~\cite{Local}.  I summarize this argument with a proposition.

\begin{proposition} 
A shape $T$ admits an isohedral tiling 
if and only if $T$ has a surround $\mathcal{S}=\{T_1,\ldots,T_n\}$
in which every $T_i$ is extendable (in which case every tile in 
the tiling is surrounded by a congruent copy of $\mathcal{S}$).
\end{proposition}


\section{SAT formulation}
\label{sec:comp}

In previous work I showed how to use a
SAT solver to compute Heesch numbers of simple polyforms~\cite{Kaplan}.
My software constructs a sequence of Boolean formulas
equivalent to the questions ``Can $T$ be surrounded at least once?'',
``Can $T$ be surrounded at least twice?'', and so on, and passes
them to a SAT solver.  It halts as soon as one of these questions
is false (or after a predetermined maximum number of levels,
to avoid looping forever when given a shape that tiles).

Here I show that it is possible to incorporate the mathematical
ideas of the previous section into my Heesch number computation,
by interposing the question ``Can $T$ tile isohedrally?'' immediately
after ``Can $T$ be surrounded at least once?''.  Indeed, the new
question is a simple restriction of the surroundability formula
already being used, taking the form ``Can~$T$ be surrounded at least
once, in a way that witnesses its ability to tile isohedrally?''.

Let $\mathcal{T}$ be a tiling of the plane.  A
\emph{poly-$\mathcal{T}$-tile} is a shape created by gluing together
a finite connected set of tiles from $\mathcal{T}$. 
Informally, I refer to a poly-$\mathcal{T}$-tile as a ``polyform'', to 
$\mathcal{T}$ as ``the grid'', and to the tiles of $\mathcal{T}$ as
``cells''.  In any patch or tiling by a polyform, I will
also require that every tile be a union of cells from the grid; that
is, every tile must be ``aligned'' to the grid.

Let $T$ be a poly-$\mathcal{T}$-tile.
Define the \emph{halo} of $T$ to be all grid cells
not in $T$ that are neighbours of cells in $T$.  Compute the 
set $\{T_1,\ldots,T_n\}$ of all transformed copies of $T$ that can
be neighbours of $T$ in a surround.  Each $T_i$ will have the
form $g_i(T)$ for a rigid motion $g_i$.  Any legal surround
must be a subset of the $T_i$ that collectively occupy every halo cell
without overlapping each other.  We can express these criteria using
a Boolean formula, a simplified version of the one I used for Heesch
number computation.  Abusing notation slightly, create Boolean
variables $T_1,\ldots,T_n$ for each potential member of the surround.
Now construct a formula with the following clauses:

\begin{itemize}
\item For every cell in the halo, a conjunction of all the $T_i$ that
use that cell (every cell in the halo must be occupied);
\item For every pair $T_i$ and $T_j$ that overlap in one or more cells,
a clause of the form $(\neg T_i \vee \neg T_j)$ (overlapping tiles
are mutually exclusive).
\end{itemize}

If a satisfying assignment is found for this formula, then a candidate
surround will correspond to the subset of variables set to true.
It is possible, however, for the resulting set of tiles to enclose
holes; if a hole is detected, then a clause is added to suppress
this solution and the SAT solver is restarted. This process iterates
until either a simply connected solution is found, or no more
candidate surrounds remain. 

If $T$ is surroundable, we can check whether it tiles isohedrally
before trying to surround it with more layers.  I do so by augmenting
the formula above with new clauses.  Let $T_i=g_i(T)$ and $T_j=g_j(T)$
be two neighbours of $T$ that are also themselves neighbours.  If
$T_i$ and $T_j$ are used together in a surround $\mathcal{S}$, then
they must both be extendable by that surround.  Note that $g_i(T_j)
= g_i\circ g_j(T)$ will be one of the shapes in $\mathcal{S}_i$,
the copy of $\mathcal{S}$ surrounding $T_i$, and must therefore
avoid conflicts with the shapes in $\mathcal{S}$.  We can enforce
this condition by finding the member $T_k=g_i\circ g_j(T)$, if it
exists, and adding a clause of the form $(\neg T_i\vee \neg T_j\vee
T_k)$ (if $T_i$ and $T_j$ are both used in a surround, then $T_k$
must be used too).  By symmetry, we perform the same steps for 
$g_j\circ g_i$.

We can add clauses to this formula that further restrict the space
of possible solutions the SAT solver must explore, potentially
improving performance.  Suppose $T_i=g_i(T)$ is part of an isohedral
surround, and $g_i$ is not an involution.  Then because $T$ is a
neighbour of $g_i(T)$, it follows that $g_i^{-1}(T)$ is a neighbour
of $T$, meaning that it must also appear in the surround.  We
therefore find $T_k = g_i^{-1}(T)$ and add a clause of the form
$(\neg T_i\vee T_k)$, which forces $T_k$ to be used if $T_i$ is.
Similarly, in the joint cases above we also add clauses for $g_i\circ
g_j^{-1}$ and $g_j\circ g_i^{-1}$, if those transformations correspond
to neighbours of $T$.

This augmented formula has a satisfying assignment if and only if 
it corresponds to a surround of $T$ for which every $T_i$ in the 
surround is extendable, or in other words, if and only if $T$
tiles the plane isohedrally.

\section{Discussion}
\label{sec:conclusion}

I implemented the augmented Boolean formula described above within
the framework of my existing software for computing Heesch numbers
of polyforms~\cite{Kaplan}.
In my implementation, transformed copies of a polyform
$T$ are represented via their affine transformation matrices (and not
their boundaries or cells).  A matrix effectively also serves as an 
asymmetric marker, thereby preventing any issues from arising
with symmetric shapes.

As a simple validation, my software produces counts of isohedral
polyforms that agree with the figures tabulated by Myers~\cite{Myers}, 
up to the size limits I tested (12-ominoes, 12-hexes, 13-iamonds, and 
12-kites).

When resigning oneself to the black box of a SAT solver, questions
of asymptotic complexity become largely moot.  Therefore, a theoretical
comparison with, say, the quasilinear-time time algorithm of Langerman
and Winslow~\cite{WL} is not particularly meaningful.  My approach
is slower than what would be possible with an efficient
implementation of their algorithm, and is certainly slower than
Myers's lightning-fast hand-optimized C code.  In the context of
my software, the extra time required for checking isohedral tilability
as part of computing Heesch numbers is minimal.  Furthermore, this
approach is remarkably convenient---the original program for computing
Heesch numbers required a few thousand lines of C++ code, and fewer
than 100 lines were added for this enhancement.  It is
also quite general: it adapts seamlessly to arbitrary polyform grids,
and does not require any special-purpose code for different isohedral
tiling types (in fact, it uses the definition of isohedral tiling
directly, and does not rely on any information about tiling types at all).

My enhanced implementation still cannot resolve the tiling-theoretic
status of every
polyform.  In particular, it is
unable to compute the isohedral number of any $k$-anisohedral
polyform (which admits only tilings containing at least $k$
transitivity classes of tile) for $k\ge 2$.  It would be interesting to 
explore
further methods based on discrete optimization that can expand to
cover these more complex, but equally important shapes.  And of course,
no software can currently detect aperiodic monotiles, for which no general
procedures are known.

\section*{Acknowledgements}
 
Thanks to Joseph Myers and Doris Schattschneider for helpful feedback
during the course of this work and the preparation of this paper.

\nocite{*}
\bibliographystyle{eptcs}
\bibliography{tilings}
\end{document}